# $T_c$=21K in epitaxial FeSe$_{0.5}$Te$_{0.5}$ thin films with biaxial compressive strain

E.Bellingeri[a], I.Pallecchi[a], R.Buzio[a], A.Gerbi[a], D.Marrè[a,b], M.R.Cimberle[c], M.Tropeano[a,b], M.Putti[a,b] A. Palenzona[a,d], C.Ferdeghini[a]

[a] CNR/INFM-LAMIA corso Perrone 24, 16152 Genova, Italy

[b] Dipartimento di Fisica, Via Dodecaneso 33, 16146 Genova, Italy

[c] CNR-IMEM, Dipartimento di Fisica, Via Dodecaneso 33, 16146 Genova, Italy

[d] Dipartimento di Chimica e Chimica Industriale, Via Dodecaneso 31, 16146 Genova, Italy

E-mail: emilio.bellingeri@lamia.infm.it

Abstract

High purity epitaxial FeSe$_{0.5}$Te$_{0.5}$ thin films with different thickness were grown by Pulsed Laser Ablation on different substrates. By varying the film thickness, $T_c$ up to 21K were observed, significantly larger than the bulk value. Structural analyses indicated that the a axis changes significantly with the film thickness and is linearly related to the $T_c$. The latter result indicates the important role of the compressive strain in enhancing $T_c$. $T_c$ is also related to both the Fe-(Se,Te) bond length and angle, suggesting the possibility of further enhancement.

Since the discovery of the iron based superconductors [1], great efforts have been devoted to the preparation of thin films of the various phases [2-7] and references therein. Between other reasons, the interest in films deposition is motivated by the strong T$_c$ dependence on external and chemical pressure in iron based pnictides and calcogenides [8-12], which has suggested the idea of exploring whether a similar effect can be induced by strain in thin films. Indeed, such expectation turned out to be true: in Ba(Fe$_{1-x}$Co$_x$)$_2$As$_2$ thin films [7] deposited on various substrates, T$_c$ has been observed to increase with the ratio c/a. Also in FeSe$_{0.5}$Te$_{0.5}$ thin films an increase of T$_c$ has been obtained by two groups [5,6] in particular a maximum T$_c$ of 17K has been measured. Such increase has been attributed to the observed c axis decrease with respect to the bulk value.

Here we present preparation as well as structural, morphological and physical characterization of epitaxial FeSe$_{0.5}$Te$_{0.5}$ thin films with different thickness, deposited on different substrates. A maximum T$_c$ =21K was obtained, which is a remarkable 30% increase with respect to the bulk value.

The films were grown by Pulsed Laser Ablation Deposition (PLD) in ultra high vacuum system [13] starting from a FeSe$_{0.5}$Te$_{0.5}$ (Fe(Se,Te)) target prepared by direct synthesis from high purity materials (Fe 99.9+%, Se 99.9% and Te 99.999%) [5].

With the aim of introducing biaxial strain we deposited films on single crystal substrates with different cell parameters, namely magnesium oxide (MgO *a*=4.217 Å) , strontium titanate (STO *a*=3.905 Å ), lantanum aluminate (LAO *a*=3.789 Å), and yttria stabilized zirconia (ZrO:Y, *a*=3.637 Å); for all the substrates we used the (*001*) orientation. The deposition conditions were optimized as reported in a previous paper [5]; namely, we used a deposition temperature of 550°C at a pressure of 5·10$^-$

$^9$ mBar. The laser repetition rate was 3 Hz (248 nm wavelength) and the laser fluency was 2 J/cm$^2$ (2 mm$^2$ spot size). The target-substrate distance was maintained at 5 cm.

In order to study the residual strain behaviour, films of different thickness from 1.2 nm to 600 nm were deposited; the thickness was calibrated by X-ray reflectometry. XRD analysis allowed to identify the PbO-like tetragonal phase for FeSe$_{0.5}$Te$_{0.5}$ grown on all the substrates, with any traces of neither elementary oxides nor hexagonal phase. In fig. 1 X-ray diffraction measurements of a film deposited on LAO are reported: in the θ-2θ scans, (fig. 1a) only the (*00l*) reflections of the film and the substrate are present, indicating the excellent purity of the phase and the optimum c-axis alignment of the growth. Typical omega scans, shown in fig 1.c, on these reflections show rocking curves having FWHM in the range 0.1° - 0.15°. The *c*-axis values extracted from the (00*l*) peak positions are in the range 5.84-5.89 Å, much smaller than the bulk value (6.03 Å), but in agreement with previous reports [4,5,6]. In order to estimate the in plane cell parameter *a*, θ-2θ scans along the reciprocal space direction [*l0l*] were performed (see fig 1b). Fitting the (*l0l*) position, we calculated the interplanar distance d$_{l0l}$ and thus the *a* axis value using the relationship $\frac{1}{d_{l0l}^2} = \frac{1}{a^2} + \frac{1}{c^2}$. The φ scan of the (*101*) reflection with a FWHM of 0.56° (fig. 1.d) indicates the high in-plane quality of the epitaxy. Film deposited on STO present similar values of rocking curve and φ scans FWHM [5], whereas broad peaks are observed on MgO indicating a poorer quality. Double-in-plane orientation tilted by a few degrees are observed in films deposited on ZrO:Y.

The $c$ and $a$ cell parameters were measured for films on LAO, STO and ZrO:Y of different thickness. Whereas no significant and systematic trends are observed for the $c$ axis, a strong correlation of the value of the in-plane $a$ axis with the thickness exists. Surprisingly, the substrate cell parameter seems to have minor or no effect at all. As shown in fig 2.a the $a$ axis value clearly decreases with increasing thickness up to approximately 200 nm (for all the substrates) and slightly increases for higher thickness. Films whose thickness was larger than 100 nm did not stick to the substrate, unless LAO substrates were employed.

This thickness effect is clearly different from what expected on the basis of simple matching between film and substrate in-plane cell parameters, i.e. when a thin film grows on a substrate with smaller lattice-parameter a large compressive strain is expected and strain relaxation occurs with increasing substrate cell parameter and/or film thickness. Conversely, in this case, the behaviour of the $a$ axis can be accounted for by considering the growth mode of our films. Indeed, at the very initial stages of the growth the RHEED images revealed a three-dimensional (transmission) diffraction pattern, which turned into an almost entirely two dimensional (surface) diffraction pattern with increasing thickness. This behaviour is confirmed by the topography observed in the AFM images of films deposited on LAO: at the beginning of the growth Fe(Se,Te) nucleates in form of isolated islands (bright spots visible on the substrate terraces) (fig. 3.a); increasing the deposition time the island concentrate at the step edges (fig. 3.b) and for a thickness of about 30 nm the islands start to coalesce (fig. 3.c) and finally an uniform substrate coverage is obtained (fig. 3.d). This behaviour is described as Volmer-Weber growth and is typical of situations where the film does not wet the substrate. In this growth mode, the coalescence may induce a large tensile strain [14-16] and reference therein. In the past, this effect was used to obtain MgB$_2$ thin films with enhanced critical temperature [17] thanks to the effect of softening of the E$_{2g}$ phonon mode by tensile strain. If the growth is carried on further, in the post coalescence regime, the tensile strain decreases and may even change in sign, becoming compressive.

In the growth process of our films, in agreement with the AFM images, the Fe(Se,Te) islands coalesce at a thickness of the order of 30 nm and, with further increase of the thickness, a biaxial compressive strain builds up, resulting in a reduced cell parameter *a*.

In fig. 3.c, resistivity versus temperature from 5K up to 300K for thin films deposited on LAO with different thickness from 9 nm up to 420 nm is shown. There is a striking evidence that the film thickness strongly affects transport and superconducting properties. In the normal state the room temperature resistivity decreases monotonically with increasing thickness, showing a crossover from semiconducting to metallic behaviour. Not surprisingly, the threshold thickness for the onset of metallic behavior is 36 nm, close to the value of island coalescence. Just at this thickness value, the superconducting transition appears as well. T$_c$ increases with thickness, up to 21K for a thickness of 200 nm. This is evidenced in fig. 3.b where T$_c$ versus thickness is shown.

In order to emphasize the correlation between the superconducting critical temperature and the strain s, we plot T$_c$ versus *a* in fig. 4.a; in the same figure the residual strain ( calculated as $\frac{a-a_0}{a_0}$, where $a_0$ is the bulk value) is also reported as upper horizontal axis. Despite the scattering of the data, due to several factors that may affect T$_c$, the two measured quantities are clearly linearly related in the considered range. We conclude that the strain induced by the growth acts in thin films as the external pressure does in bulk samples [18], yielding a significant enhancement of the critical temperature.

With the aim of further clarifying the structural effect of the strain and its correlation with the superconducting properties we estimated the vertical position (*z*) of the (Se,Te) atoms inside the cell. We simulated the XRD spectra of fully *c* axis oriented films with different *z* and we found that the intensity of the (*003*) reflection is strongly affected by this value and not much affected by other parameters. It was thus possible to refine confidently the *z* position (in the P4/nmm space group) on the symmetric scans diffraction data, using the FULLPROF software [19]. Due to the superposition of the STO (*002*) peak with the Fe(Se,Te) (*003*), this data analysis was possible only for the samples on LAO. Finally, from the *a* , *c* and *z* values, we calculated the length (*d*) and the angle (α) of the Fe-(Se,Te) bond; the results are reported in fig. 4.b and 4.c respectively. These two structural parameters are often indicated in literature as the key ones that control superconductivity in oxipnictides and calchogenides, as they determine the exchange interaction responsible for the Fe spin ordering [20-22]. A maximum T$_c$ is expected when the bond angle approaches the value 109.47° (indicated by a vertical dashed line in figure 4.c) , which corresponds to a regular tetrahedron, and the bond length approaches the ideal value of 2.40 Å. In our case this expected trend is confirmed: the samples with higher T$_c$ are just the ones whose bond angles and lengths approach the respective ideal values.

In summary, we grew high purity epitaxial FeSe$_{0.5}$Te$_{0.5}$ thin films by PLD deposition. STO and LAO resulted to be the best substrates, on which fully epitaxial growth was obtained. On LAO it was possible to deposit films up to 400 nm. A remarkable T$_c$ increase was obtained in 200 nm thick films (T$_c$= 21K to be compared with 16.2 K, bulk value). In addition an unambiguous correlation between T$_c$ and the *a*-axis was found, which implies a compressive strain driving such increase. The decrease of *a* is accompanied by a decrease of the Fe-(Se,Te) distance and by an increase of the (Se,Te)-Fe-(Se,Te) bond angle α toward the values which are thought to be the optimal ones for superconductivity. As long as such optimal values are not achieved yet, there may be still and edge of improvement of Tc in epitaxial FeSe$_{0.5}$Te$_{0.5}$ thin films.

Acknowledgments

The financial support of Fondazione San Paolo is acknowledged.

**Figure captions**

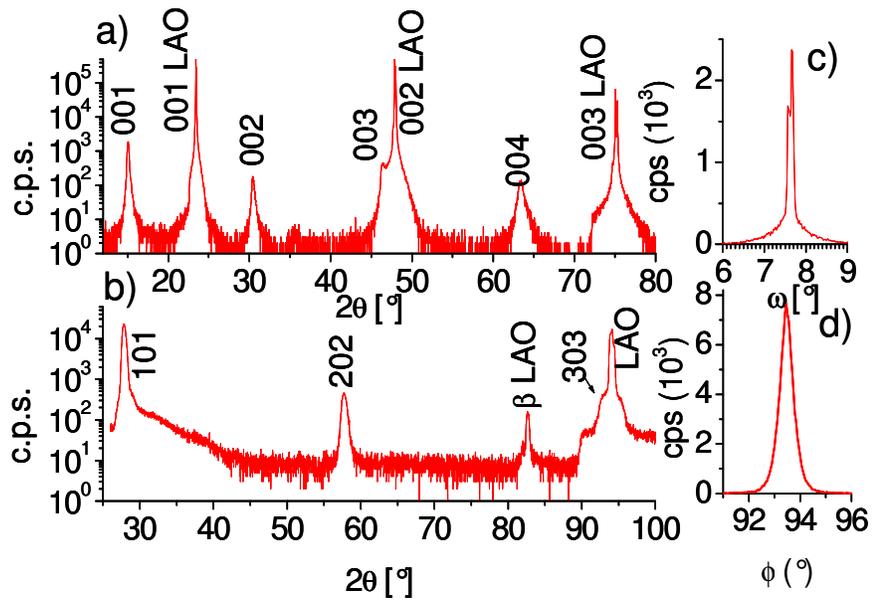

Fig. 1: x-ray diffraction data on Fe(Se,Te) thin film deposited LAO: in a) the symmetrical θ–2θ scan, in b) θ–2θ scans in the *l0l* direction (χ=57.7°). The out of plane and in plane epitaxial quality of the growth is shown in panels c) and d) where a ω scan (rocking curve) of the 001 reflection and a φ scan of the 101 reflection are plotted, respectively.

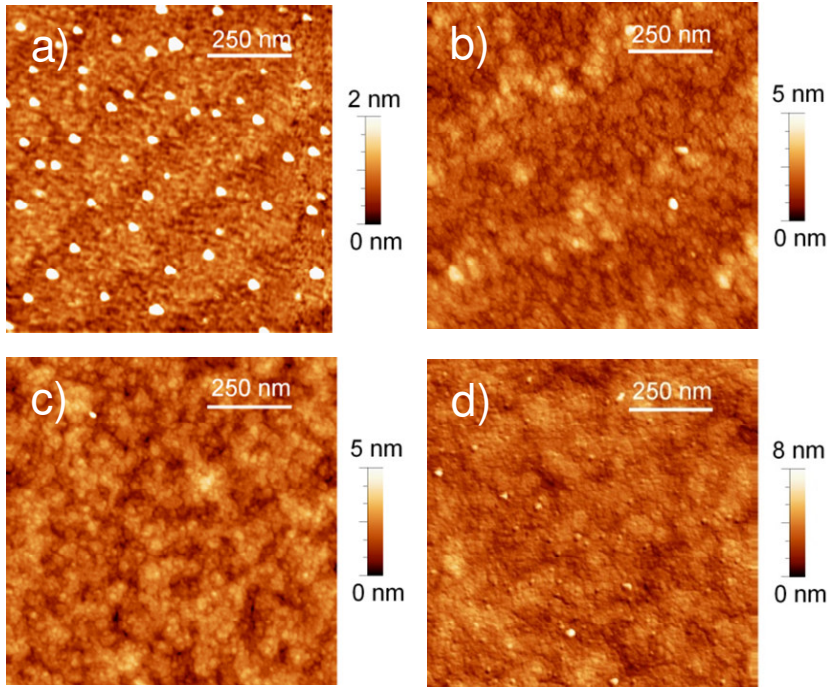

Fig 2: Atomic force microscope images of the initial and intermediate stages of the growth of a Fe(Se,Te) thin film on LAO substrate. Fe(Se,Te) initially nucleates in the form of small islands a) that concentrate at the step edges b); the islands coalesce c) and a uniform coverage is obtained d).

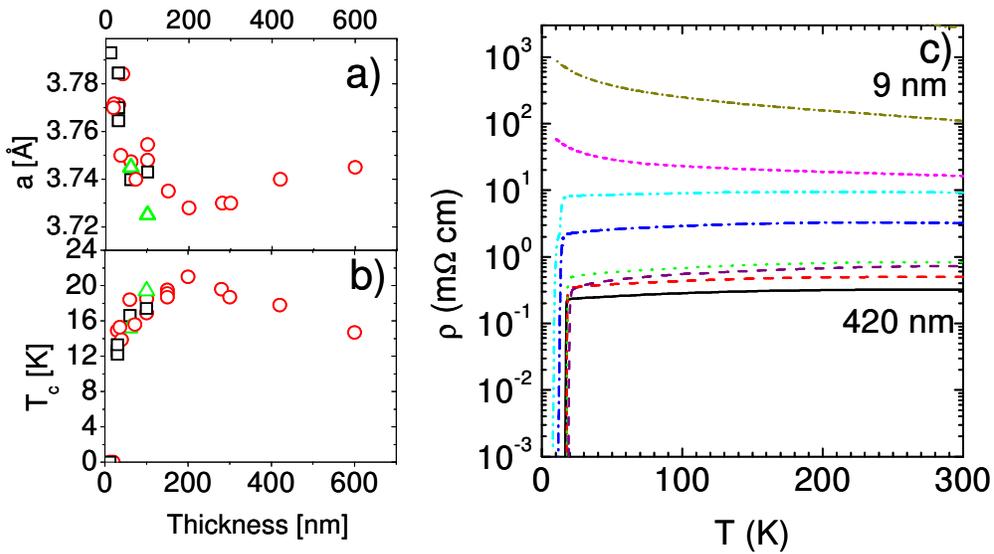

Fig. 3: In panel a) the resistivity of several Fe(Se,Te) films with different thickness (9, 18, 36, 72, 150, 200, 280 and 420 nm) are reported: the room temperature resistivity monotonically decreases increasing the thickness from 9 to 420 nm, whereas the superconducting critical temperature, (panel b) reaches a maximum value of 21K for a thickness of about 200 nm; for the same thickness value a minimum value of the in-plane $a$ cell parameter is observed (panel c). In b) and c) in addition to samples deposited on LAO (○) T$_c$ and $a$ values measured on film deposited on STO (□) and ZrO:Y (△) are reported.

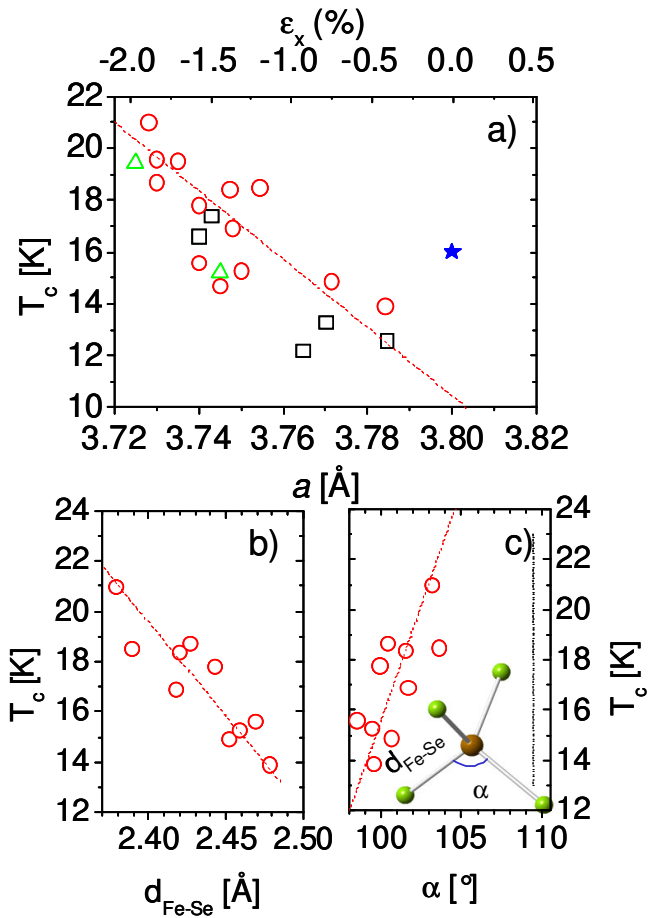

Fig. 4: In panel a) the transition temperatures of several Fe(Se,Te) films on LAO (O) STO (□) and Y:ZrO (Δ) are reported as a function of the in-plane cell parameter $a$; the star represents the bulk value. In panels b) and c) the transition temperatures of films deposited on LaAlO3 are plotted as a function of the Fe-(Se,Te) bond length and of the (Se,Te)-Fe-(Se,Te) bond angle, respectively. As an eye guideline the linear fit of the data (dashed lines) is added in all the panels.